\documentclass[aps,preprint,showpacs,showkeys]{revtex4}%
\usepackage{amsfonts}
\usepackage{amsmath}
\usepackage{amssymb}
\usepackage{graphicx}%
\providecommand{\U}[1]{\protect\rule{.1in}{.1in}}

\def\lsim{\mathrel{\raise.3ex\hbox{$<$\kern-.75em\lower1ex\hbox{$\sim$}}}}
\begin{document}
\title[Breaking of Conformal Symmetry in a Superpotential]{Radiatively Induced Breaking of Conformal Symmetry in a Superpotential}
\author{A. B. Arbuzov}
\affiliation{Bogoliubov Laboratory of Theoretical Physics, Joint Institute for Nuclear
Research, 141980 Dubna, Russia}
\affiliation{Department of Higher Mathematics, Dubna State University, 141982 Dubna, Russia}
\author{D. J. Cirilo-Lombardo}
\affiliation{Bogoliubov Laboratory of Theoretical Physics, Joint Institute for Nuclear
Research, 141980 Dubna, Russia }
\affiliation{National Institute of Plasma Physics(INFIP-CONICET), FCEyN, Department of
Physics, Universidad de Buenos Aires, Buenos Aires 1428, Argentina}
\email{diego777jcl@gmail.com}

\begin{abstract}
Radiatively induced symmetry breaking is considered for a toy model with one
scalar and one fermion field unified in a superfield. It is shown that the
classical quartic self-interaction of the superfield possesses a quantum
infrared singularity. Application of the Coleman-Weinberg mechanism for
effective potential leads to the appearance of condensates and masses for both
scalar and fermion components. That induces a spontaneous breaking of the
initial classical symmetries: the supersymmetry and the conformal one. The
energy scales for the scalar and fermion condensates appear to be of the same
order, while the renormalization scale is many orders of magnitude higher. A
possibility to relate the considered toy model to conformal symmetry breaking
in the Standard Model is discussed.

\end{abstract}
\volumeyear{year}
\volumenumber{number}
\issuenumber{number}
\eid{identifier}
\date[Date text]{date}
\received[Received text]{date}

\revised[Revised text]{date}

\accepted[Accepted text]{date}

\published[Published text]{date}

\maketitle




\section{Introduction}

It was shown~\cite{Coleman:1973jx} that infrared divergences in quantum loop
contributions to effective potentials of various models in quantum field
theory (QFT) can lead to the necessity to introduce a non-zero renormalization
scale and thus generate a spontaneous breaking of the conformal symmetry. In
particular, the $\phi^{4}$ model as well as Abelian and Yang-Mills gauge
models were considered in~\cite{Coleman:1973jx}. Here we will apply the
Coleman--Weinberg (CW) mechanism to a simple QFT model for a superfield which
joins scalar and fermion physical fields, see for example
\cite{Edery:2013dqa,Edery:2015zda} for application of the CW mechanism in
different physical scenarios.

From the phenomenological point of view our study is motivated by the recent
discovery of the Higgs boson. The observed properties of the latter are in a
good agreement with the Standard Model (SM) predictions. Nevertheless the
origin of the electroweak energy scale is still unclear. For the time being it
is just introduced from the beginning into the Lagrangian of the SM as the
tachyon-like mass parameter. On the other hand the electroweak (EW) energy
scale of about 100~GeV is seen both in the Higgs (and electroweak sector) and
the top quark mass. The relation $4m^{2}_{H} \approx2m^{2}_{t} \approx v^{2}$
between the observed Higgs boson mass $m_{H}$, the top quark mass $m_{t}$, and
the Higgs boson vacuum expectation value $v$ holds with a high
accuracy~\cite{Gorsky:2014una}. In any case, the coincidence of the scales is
an intriguing puzzle. Another face of the electroweak scale puzzle is the
hierarchy problem of the SM due to quadratic divergences in the running of the
Higgs boson mass within the SM. In fact, renormalization of $m_{H}$ suffers
from \emph{fine tuning} between the (loop) contributions due to the top quark,
the Higgs boson, and EW bosons (note that only longitudinal components of $W$
and $Z$ bosons, \textit{i.e.} the scalar Goldstones, contribute). It is well
known that resolution of the fine tuning problem can be done by a
supersymmetric extension of the SM. In any case, a certain (symmetry?)
relation between fermionic and bosonic contribution is required to solve the problem.

On the other hand there are many indirect indications the the Conformal
Symmetry (CS) might be the proper symmetry of the underlying fundamental
theory, while the SM is just an effective model emerged after a spontaneous
breaking of the CS, see e.g.
Refs.~\cite{Chankowski:2014fva,Heikinheimo:2013fta}.

In Ref.~\cite{Arbuzov:2016xte}, the possibility to generate a soft breaking of
the conformal symmetry in the sector of the SM which joins the Higgs boson and
the top quark was discussed. In fact, the infrared singularity is present in
this system and the Coleman--Weinberg mechanism can be applied. Nevertheless,
the question about the relation between renormalization conditions for the
scalar and the fermion field remains unsolved. As discussed, a certain
\emph{bootstrap} should happen in the SM between the Higgs boson and the top
quark. In this paper we suggest to look what comes out if the two fields are
joined into a superfield. Of course, it is just one of many other
possibilities but it provides a certain feeling of the bootstrap.

It is frequent to find in the current research proposals that suggest
(supersymmetric) extensions of the Standard Model with introduction of a
hidden sector that shows conformal invariance above a certain high energy
scale (see in a different context Ref.~\cite{Heikinheimo:2013xua}), and it
couples to the SM sector by some higher dimensional operators. The nontrivial
conformally invariant hidden sector leads to a novel type of observable
effects in the SM sector, which may be accessible in near future experiments
at a TeV scale. On the other hand, since one of the most appealing new physics
at the TeV scale is the supersymmetry (SUSY), it is very natural to consider
supersymmetric extension of the Standard Model considering the introduction of
superfields containing the particles involved in the interaction of interest.
The first aim of this paper is to investigate the supersymmetric extension of
the model based on the superconformal field theory by means of introduction of
a scalar superfield, as we will show in Section II. It is well known that the
four dimensional superconformal field theory is powerful enough to obtain the
crucial dynamical information about the physics of particle interaction due to
the fact that the interaction itself is hidden inside quadratic and dynamical
terms in the Lagrangian of the theory. For example, the relation between the
R-charge and the conformal dimension determines the conformal dimensions of
the chiral operators beyond the perturbation theory. We also have more severe
inequalities for conformal dimensions that are not available in
non-supersymmetric theories. In this sense, the introduction of the SUSY is
theoretically well motivated. Physically the interplay between conformal
symmetry and supersymmetry and their breaking (of both or of any of them,
total or partial) introduces automatically extra constraints on particle physics.

As we pointed out before, previous investigations on the particle physics
within the context of (super) conformal models assume that a certain particle
sector remains conformal at least down to the electroweak scale, at which any
experimental evidence is expected. The problem is claimed usually of a partial
breaking of SUSY or conformal symmetry and how to conciliate both. It is
usually solved by means of a gauge mediation or by tuning the K\"{a}hler
potential. However there are no such problems through our paper, consequently
this particular point not will be analyzed here.

\section{Conformal Models and Supersymmetry}

The hints in order to introduce the fermionic interactions in any classical
bosonic action endowed by conformal symmetry were presented for the first time
in the seminal papers of Akulov and Pashnev~\cite{Akulov} where the starting
point was the well know the AFF\ (de Alfaro, Fubini and Furlan) conformal
model~\cite{deAlfaro}. Without going into details (see~\cite{Akulov}), the
idea was to introduce a superfield having the form
\begin{equation}
\Phi=\varphi+i\theta^{\alpha}\psi_{\alpha}+i\theta^{\alpha}\theta_{\alpha
}F\text{ ,\ \ \ \ (}\theta^{\alpha}\theta_{\alpha}=\theta\overline{\theta
}\text{ etc.)} \label{Phi}%
\end{equation}
into the following general n-dimensional action with the standard
super-kinetic term
\begin{equation}
S_{kin}=-\frac{1}{32}\int d^{n}xd^{2}\theta\text{ }D^{2}\overline{D}%
^{2}\left(  \overline{\Phi}\Phi\right)  \label{s-kin}%
\end{equation}
where $\Phi$ is the chiral superfield (with standard anti-chiral counterpart
$\overline{\Phi}$ ) and the super-derivatives are usually defined as
$D_{\alpha}\Phi\equiv\left(  \frac{\partial}{\partial\theta^{\alpha}}+i\left(
b_{\alpha\beta}\theta^{\beta}\right)  ^{i}\frac{\partial}{\partial x^{i}%
}\right)  \Phi$ (similarly for$\overline{D}_{\overset{\cdot}{\alpha}}%
\overline{\Phi}$ )where $b_{\alpha\beta}$ is a symmetric matrix fixed by the
symmetry properties of the superspace under consideration, e.g. by
supercharges. The usual conventions for down and up indices of the fermionic
variables with $\epsilon^{12}=\epsilon_{12}=1$, $\left( \alpha,\beta
=1,2\right) $ are assumed (for the dotted indices $\overset{\cdot}{\alpha
},\overset{\cdot}{\beta}=\overset{\cdot}{1},\overset{\cdot}{2}$ are similarly
related, as usual$)$, also for spacetime indices: $i,j=0,....,d-1$. The
component form of expression (\ref{s-kin}) is obtained by inserting
(\ref{Phi}) into (\ref{s-kin}) and integrating over the Grassman variables:
\begin{equation}
S_{kin} = -\frac{1}{2}\int d^{n}x\text{ }\left(  \partial_{i}\varphi
\partial^{i}\overline{\varphi} -i\left(  \overline{\psi}^{\overset{\cdot
}{\alpha}} b_{\overset{\cdot}{\alpha}\beta}\right) _{j}\partial^{j}\psi
^{\beta} +4F\overline{F}\right) .
\end{equation}
The interaction part was defined in the general form as
\begin{equation}
S_{int}=\int d^{n}xd^{2}\theta\,d^{2}\overline{\theta}V(\Phi\overline{\Phi}).
\end{equation}

Without loss of generality the simplest 4-dimensional case will be treated.
Remind now the effective potential for a scalar field with a $\varphi^{4}$
interaction, which was derived by S.~Coleman and
E.~Weinberg~\cite{Coleman:1973jx} in the one-loop approximation
\begin{equation}
U(\varphi)=\frac{\lambda}{4!}\varphi^{4}+\frac{\lambda^{2}}{256\pi^{2}}%
\varphi^{4}\left[  \ln\left(  \frac{\varphi^{2}}{M^{2}}\right)  -\frac{25}%
{6}\right]  .
\end{equation}
The presence of the renormalization scale $M$ indicates the radiatively
induced breaking of the conformal symmetry in this model.

We can pass from the bosonic effective potential to the supersymmetric one by
introducing the superfield. Note that due that the standard version of the
four-dimensional supersymmetry, the simplest superfields contain a complex
Lorentz scalar and a chiral (left-handed or right-handed) fermion. To avoid
confusion henceforth we define $\left\langle \varphi\right\rangle ^{2}%
\equiv\overline{\varphi}\varphi$. Then we obtain the following expression
\begin{align}
W(\left\langle \varphi\right\rangle ,\left\langle \bar{\psi}\psi\right\rangle
)  & = \left(  \left\langle \varphi\right\rangle ^{4}+2\left\langle
\varphi\right\rangle \langle\bar{\psi}\psi\rangle\right)  \left[
\frac{\lambda}{4!}+\frac{\lambda^{2}}{256\pi^{2}}\left(  \ln\left(
\frac{\left\langle \varphi\right\rangle ^{2}}{M^{2}}\right)  -\frac{25}%
{6}\right)  \right] \nonumber\\
& + \left\langle \varphi\right\rangle \langle\bar{\psi}\psi\rangle\left[
\frac{\lambda}{2}+\frac{\lambda^{2}}{256\pi^{2}}\right] ,
\end{align}
where the Grassman integration was performed under the (physical) measure
\begin{align}
&  \int\mu\left(  \theta^{2}\right)  d^{2}\theta=b \quad\text{ and } \int
\mu\left(  \theta^{2}\right)  \theta^{2}d^{2}\theta=a\nonumber\\
&  \text{with }\ \mu\left(  \theta^{2}\right)  \equiv a\exp\left(
b\frac{\theta^{2}}{a}\right) ,\nonumber
\end{align}
where $a$ and $b$ are constants related to the group manifold structure
(volume), that must be included into the above measure in order to recover the
original Coleman--Weinberg potential when all fermions vanish.

Let us look for a minimum of the potential. The conditions
\begin{align}
\left\{
\begin{array}
[c]{l}%
\frac{\partial W}{\partial\langle\bar{\psi}\psi\rangle} = 0,\\
\frac{\partial W}{\partial\langle\varphi\rangle} = 0
\end{array}
\right.
\end{align}
lead to the following solution for the scalar and fermion condensate values:
\begin{align}
\label{Mscale} &  v^{2} \equiv\langle\varphi\rangle^{2} = M^{2} \exp\left\{
-\frac{196\pi^{2}}{\lambda}\right\}  ,\nonumber\\
&  \langle\bar{\psi}\psi\rangle= - v^{3}\frac{2\lambda}{7}\, .
\end{align}
We assumed that the coupling constant $\lambda
\mathrel{\raise.3ex\hbox{$<$\kern-.75em\lower1ex\hbox{$\sim$}}} 1$ so that the
perturbative solution is reliable.

We would like to make a parallel to the sector of the Standard Model, which
joins the Higgs self-interaction and the Yukawa term of the top quark,
see~\cite{Arbuzov:2016xte} for details. In fact the structure of this sector
is exactly the same as the one of our toy model. The condition $\lambda
\mathrel{\raise.3ex\hbox{$<$\kern-.75em\lower1ex\hbox{$\sim$}}}1$ holds in the
SM. Taking realistic SM values of $\lambda$ and $v\approx246$~GeV, we see that
the scale $M$ appears to be extremely large: $M\gg M_{\mathrm{Planck}}$. This
value emerged in our toy model, but the general hierarchy between the EW scale
and the renormalization scale $M$ does naturally appear in the
Coleman--Weinberg mechanism applied to a model of interacting scalar and
fermion particles with any assumption on a symmetry relation between these two
fields. This allows us to speculate about the possibility to have the Planck
scale as the proper renormalization scale of the Standard Model being
responsible for the scale invariance breaking. The source of the large
difference between to EW scale and the Planck mass can be provided just by the
exponent in a relation similar to Eq.~(\ref{Mscale}).

The spontaneous breaking of the conformal symmetry in the system leads to
generation of masses both for the scalar and fermion fields in the standard
way after the shift of the scalar field $\phi=h+v$:
\begin{equation}
m_{h}=\sqrt{\lambda}\frac{v}{\sqrt{2}},\qquad m_{f}=\frac{7}{12}\lambda v.
\end{equation}
Note that the energy scale as for the masses as well as for the condensates of
both fields is the same:
\begin{equation}
m_{h}\sim m_{f}\sim v\sim-\sqrt[3]{\langle\bar{\psi}\psi\rangle}.
\end{equation}
Remind that the coincidence of scales of the Higgs boson mass and of the top
quark one is one of the puzzles in the SM.

In the case if we take the model with a K\"{a}hler structure, the potential is
slightly modified as:
\begin{align}
W(\left\langle \varphi\right\rangle ,\left\langle \bar{\psi}\psi\right\rangle
)  &  \approx2\left(  \left\langle \varphi\right\rangle ^{4}+2\left\langle
\varphi\right\rangle \langle\bar{\psi}\psi\rangle\right)  \left[
\frac{\lambda}{4!}+\frac{\lambda^{2}}{256\pi^{2}}\left(  \ln\left(
\frac{\left\langle \varphi\right\rangle ^{2}}{M^{2}}\right)  -\frac{25}%
{6}\right)  \right] \nonumber\\
&  +2\left\langle \varphi\right\rangle \langle\bar{\psi}\psi\rangle\left[
\frac{\lambda^{2}}{256\pi^{2}}+\lambda\ln\left(  \left\vert \frac
{c\left\langle \varphi\right\rangle \langle\bar{\psi}\psi\rangle}{r}%
{M}\right\vert \right)  \right]
\end{align}
with $\left\vert c\right\vert <1$ that for $sign\left(  c\right)  =+1$ we have
compact manifold, and for $sign\left(  c\right)  =-1$ we have a non-compact
one. The modified potential leads to a similar solution for the minimum position.

\section{Discussion}

In general, even without any (super)symmetry between the scalar and fermion
fields we should look for the minimum of the effective potential. The symmetry
condition just helps to derive a relation between the condensate values.
Certainly, the top quark and the Higgs boson in the SM are not related by
SUSY. On the other hand, we clearly see that they are somehow tightly linked
to each other. So, we take SUSY just as a toy model to test the connection
between $t$ and $H$. We suppose that some of the features which appear in the
SUSY relation might be relevant for the true (still unclear) picture. Because,
the physical states, that we are interested in, become to be part of a same
supersymmetric multiplet, they are not independent. This fact reduces the
number of independent coupling constants. Notice that the question why the
Yukawa coupling of the top quark is just one within error bars looks like a
puzzle. It could be treated as an accidental coincidence, if it would not be
so much important for the naturalness problem in the SM. The another important
issue to have into account is about the possibility to avoid the fact that in
the supersymmetric toy model there is only a single coupling constant: when
applied to the Standard Model the requirement to put the top quark and the
Higgs in the same multiplet appears to conflict (in the case of the realistic
full model) because of their different charges. This problem can be treated,
for example, by introducing an extra symmetry (complex, quaternionic or
octonionic) at the level of the fields without modifying the original symmetry
of the model supersymmetric or not~\cite{Afonso:2012zj}. Note that instead of
introducing multiple Higgs to provide partners for the remaining quarks, we
can introduce only one Higgs with quaternionic symmetry~\cite{DiegoAndrej}
(for example, in the Weinberg-Salaam model the main idea in order to increase
the number of fields is based on the observation that there exists the
following underlying quaternionic symmetry, namely $\frac{1}{2}\left(
g^{\prime}B_{\mu}+g\sigma_{i}A_{\mu}^{i}\right)  \equiv Q_{\mu}$, this issue
is now under research~\cite{DiegoAndrej}).

It is worth to note that there is a close relation between breaking of the
conformal symmetry and SUSY. It was was pointed out in
Ref.~\cite{Akulov:1986vn} that quantization of theories within the Hamiltonian
formulation suffers from difficulties associated with the ordering of
operators. Moreover, the presence of fermionic operators creates additional
difficulties that are translated to the breaking of symmetries of the physical
system under consideration. These ordering difficulties are present as in the
definition of SUSY charges as well as in the corresponding SUSY generators.
Generally, they appear as operators with arbitrary factors that take into
account all ways of ordering where in the two solutions are non-normed which
corresponds to the case of spontaneously broken supersymmetry. Thus,
spontaneous breaking of supersymmetry at the quantum level is possible due the
indefiniteness in ordering of operators.

Since normalization is not inherent to the conformal symmetry condition, it
seems apparently that the breaking of SUSY would not affect the conformal
symmetry. But at the quantum level, the ambiguity in the ordering of operators
establishes a connection between both symmetry breaking. Precisely this
condition seems to be related to what happens in the breaking of supersymmetry
at finite temperature, even in the microcanonical
picture~\cite{CiriloLombardo:2007yg}. This allows one to make a conjecture on
triality between supersymmetry breaking, breaking of conformal symmetry, and
non-zero temperature from the quantum level. Note also that since LHC started
looking for superpartners, the task becomes extremely hard. Probably the
difficulties to interpret the absence of hints for supersymmetry at LHC
implies that there exist a supersymmetric $\Lambda$ value that can be greater
than expected, consequently higher values of $\Lambda$ can be justified in
such a case.

Searching for similar effects in other quantum mechanical models is of
considerable interest and will allow further studies of the phenomenon of
spontaneous symmetry breaking in physics.


\subsection*{Acknowledgments}

We are grateful to prof. V.N.~Pervushin for fruitful discussions and to profs.
Ariel Edery and Matti Heikinheimo for put our attention on their works. One of
us (A.A.) thanks the Dynasty foundation for financial support and D.J.C-L. is
grateful to the JINR Directorate for hospitality and to CONICET (Argentina)
for financial support.

\end{document}